\documentclass[a4paper,12pt]{article}
\usepackage{arxiv}
\usepackage[utf8]{inputenc} 
\usepackage[T1]{fontenc}    
\usepackage{hyperref}       
\usepackage{url}            
\usepackage{booktabs}       
\usepackage{amsfonts}       
\usepackage{subfigure}
\usepackage{amsmath}
\usepackage{nicefrac}       
\usepackage{microtype}      
\usepackage{lipsum}		
\usepackage{graphicx}
\usepackage{natbib}
\usepackage{doi}
\usepackage{tikz}
\usetikzlibrary{positioning, arrows.meta, backgrounds}

\title{Modelling Asset Price Dynamics with Investor Inertia: Diffusion with Advection and Fourth-Order Extension}

\author{ \href{https://orcid.org/0009-0003-3560-4322}{\includegraphics[scale=0.06]{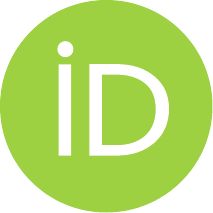}\hspace{1mm}Diego da Silva Santos}\thanks{webpage: http://lattes.cnpq.br/5030520742156057, alternative
		address: diegosantos489@gmail.com} \\
	Institute of Mathematics, Statistics and Scientific Computing-University of Campinas\\
	Campinas, BR \\
	\texttt{d215862@dac.unicamp.br} \\
	\And
	\href{https://orcid.org/0000-0002-1680-6303}{\includegraphics[scale=0.06]{orcid.pdf}\hspace{1mm}Luiz Gustavo Bastos Pinho} \\
	Department of Statistics and Applied Mathematics-Federal University of Ceará\\
	Fortaleza, BR \\
	\texttt{lgbpinho@gmail.com} \\
}

\renewcommand{\shorttitle}{\textit{}}

\hypersetup{
pdftitle={An Advection-Diffusion Model Incorporating Investor Inertia for the Dynamics of Financial Asset Prices},
pdfsubject={q-bio.NC, q-bio.QM},
pdfauthor={Diego da Silva Santos, Luiz Gustavo Bastos Pinho},
pdfkeywords={Diffusion with Advection, Financil Market, Asset Princing, Investidor Inertia},
}

\begin{document}
\maketitle

\begin{abstract}
	Standard models of asset price dynamics, such as geometric Brownian motion, do not formally incorporate investor inertia. This paper presents a two-stage framework for modelling this behaviour. First, we establish a microfoundation for the classic advection-diffusion model by representing the asset's log price as a three-state random walk (up, down or neutral). While this derivation offers a clear behavioural origin for drift and volatility, it is ultimately limited by its Gaussian nature and fails to capture the heavy tails (leptokurtosis) observed in financial markets. To address this issue, we introduce and apply a fourth-order extension inspired by diffusion with retention models, where a more complex representation of inertia generates non-Gaussian dynamics. Through an empirical application using Brazilian PETR4.SA data, we demonstrate that this extended model significantly outperforms the original in fitting the real distribution of returns. Our findings suggest that investor inertia is a dual concept capable of explaining both standard market trends and extreme events.
\end{abstract}

\keywords{Diffusion with Advection \and Diffusion with Retetion \and Financial Market \and Asset Pricing \and Investidor Inertia}

\section{Introduction}
The notion that evolution of asset prices can be characterized by stochastic processes originates from the work of \cite{bachelier1900theorie}. In his thesis ``\textit{Th\'eorie de la Sp\'eculation}'', he became the first to use Brownian motion to model market fluctuations, making a direct comparison with the random movement of particles examined in physics. Although original, the work of \cite{bachelier1900theorie} went largely unnoticed for more than fifty years. The concept was rediscovered and improved upon independently. The physicist \cite{osborne1959brownian} strengthened the analogy with physics and made an important correction by noting that it is the log-price of assets that follows a random path, not their nominal price. Soon after, \cite{samuelson2016proof} combined these concepts and provided them with a solid theoretical foundation in economics. Shortly thereafter, \cite{samuelson2016proof} synthesized these ideas, providing them with a rigorous economic theoretical foundation and establishing Geometric Brownian Motion as the standard model for capturing the unpredictable nature of markets.

Nonetheless, these models typically depend on the simplifying assumption that the asset price experiences a displacement, whether positive or negative, at every moment. This binary view does not explicitly incorporate a fundamental behavior observed in the market: investor inertia \cite{barber2000trading}, whereby periods of collective indecision result in the predominant choice being not to trade, thereby maintaining prices at their current levels.

This gap is addressed in two steps in this work.  Basically, we incorporate investor inertia from basic principles to create a microfoundation for the standard model.  We use a three-state discrete random walk to describe the log-price of an asset, allowing for three possible movements: up (buy), down (sell), or neutral (hold).  We show that this framework naturally converges to the classical advection-diffusion equation in the continuous limit.  This first contribution is important because it gives the model's main parameters distinct behavioral origins: volatility results from overall trading activity, while the advection (drift) term is caused by the asymmetry between buying and selling decisions.

Despite the fact that this derivation provides useful information, the final model is ultimately limited by its Gaussian character, which ensures normally distributed log returns.   This contradicts one of the most widely accepted stylized facts about financial markets: the presence of heavy tails, or leptokurtosis, which implies that violent price fluctuations occur far more frequently than a normal distribution would suggest \cite{mandelbrot1963variation, fama1965behavior, cont2001empirical}.

To address this constraint, we suggest and evaluate a fourth-order extension that draws inspiration from physics' diffusion-with-retention models (\cite{bevilacqua2011significance}).  In \cite{bevilacqua2015second}, he suggested a conceptual application of this 'second-order diffusion theory' to describe capital movement, demonstrating the relevance of this technique to economics as well.  He maintained that situations involving simultaneous inflows and outflows are best represented by this theory.  In our empirical implementation, we use the t-Student's distribution, a proven technique for modeling fat-tailed distributions in financial returns, to simulate the complicated dynamics produced by the fourth-order term \cite{praetz1972distribution}.

Finally, we conducted an empirical assessment of the advection-diffusion model and its fourth-order extension using historical data from the Brazilian stock market (PETR4.SA).  Because it is modeled using an t-Student's distribution, which captures the heavy tails of the log-return distribution, the results verify that the retention-diffusion equation-based model is substantially more successful than the conventional advection-diffusion model.  Furthermore, this paper provides a fresh macroeconomic interpretation of the advection-diffusion equation's volatility and drift parameters.  A three-state random walk that accounts for investor inertia naturally yields these parameters.  When used in price modeling, it also provides an interpretation of the fourth-order factor in the diffusion with retention equation, which is represented as investor inertia.

\section{Theoretical Foundations}
In this section, we will review the fundamental concepts that underpin modern asset pricing modelling. We start with the intuitive idea of a random walk, show how it converges to the diffusion equation and trace the mathematical developments that led to the standard model in finance. This foundation is essential for providing context for the introduction of diffusion model with retention.

\subsection{The diffusion equation and asset price dynamics}
A procedure known as a ``random walk'' produces random values for a variable that represents a specific location.  The probability of being at point $x$ at time $t$, represented by $p(x,t)$, changes from neighboring places in a discrete model.  Assume that the coordinates $x$, $x-a$, and $x+a$ are known.  The probability of being at place $x$ at time $t+\Delta t$ is what we wish to determine.  Two conceivable conclusions are that it was at $x-a$ or $x+a$.  Thus, we have that:
\begin{equation}
	p(x,t+\Delta t)=\alpha p(x-a,t)+\beta p(x+a,t). \nonumber
\end{equation}
That is, if we know the probability of being at the current time, then we also know the probabilities of being at a certain future time. This is because it is a Markovian process. For a symmetric walk, where the probability of taking a step up or down by a distance $a$ is the same ($\alpha = \beta = 1/2$), the relationship is:
\begin{equation}
	p(x,t+\Delta t)=\frac{p(x-a,t)+p(x+a,t)}{2}.\nonumber
\end{equation}
In the continuous limit, where the time $\Delta t$ and space $a$ intervals become infinitesimal, this recurrence relation converges to the diffusion equation:
\begin{equation}\label{eq:difusao}
	\frac{\partial p(x,t)}{\partial t}= k\frac{\partial^2 p(x,t)}{\partial x^2},
\end{equation}
where $k=\lim \frac{a^2}{2\Delta t}$, is the diffusion coefficient. This equation is used in physics to simulate how particles dissipate through a material.  It can be used, for instance, to explain the temporal evolution of a function $p(x,t)$ that expresses the temperature at a specific location at a specific time.  Another illustration is when particles in a fluid flow via a pipe, travel from regions of higher concentration to regions of lower concentration. This movement can be explained by thinking of the flow along the pipe in tiny time increments.

The first to apply this formalism to financial markets was \cite{bachelier1900theorie}. In his thesis, he postulated that the price of an asset $S$ in a ``fair game'' follows a driftless random walk. This means that price movements are unpredictable and that the expected value of the asset in the future is its current price. He concluded that the probability density of prices, $p(S,t)$, should satisfy the pure heat equation:
\begin{equation}\label{eq:difusaoS}
	\frac{\partial p(S,t)}{\partial t}=k\frac{\partial^2p(S,t)}{\partial S^2}.
\end{equation}
By analogy, the equation illustrates how unpredictable `shocks' in trade cause certainty about today's prices to diffuse into a cloud of probabilities over future values. 

After establishing the connection between the continuous diffusion equation and a discrete random walk, we move on to the formal formulation of this continuous-time random walk.  This limit is represented mathematically by the Wiener process, which is the fundamental building block for financial stochastic dynamics modeling.

\subsection{The Wiener process}
The path taken by a single `particle' (i.e. the price) in a diffusion process can be modelled mathematically using the Wiener process, also known as Brownian motion. Formally, a standard Wiener process $(X_t)_{t \geq 0}$ is a continous-time stochastic process satisfying the following properties:
\begin{itemize}
	\item[1.] $X_0=0$.
	\item[2.] For ant $0 \leq s < t$, the increment $X_t - X_s$ is independent of the history of the process up to time $s$ (Markov property).
	\item[3.] $X_t-X_s \sim N(0, t-s)$, implying zero mean and variance proportional to the elapsed time.
	\item[4.] The function $t \mapsto X_t$ is continous almost surely, but nowhere differentiable, reflecting the erratic nature of Brownian motion.
\end{itemize}

Over an infinitesimal time interval $dt$, the increment $dX_t$ of a standard Wiener process satisfies:
$$E(dX_t)=0, \mbox{ } Var(dX_t)=dt, \mbox{ } dX_t \mbox{ }\bot \mbox{ } dX_s \mbox{ } \mbox{ for} \mbox{ } s \neq t.$$

The generalised Wiener process extends this model to include deterministic trends (drift) and stochastic volatility, fundamental in financial modeling \cite{merton1971theory, duffie2010dynamic}. It is described by the stochastic differencial equation (SDE):
\begin{equation}
	dx_t=\alpha dt+\sigma dX_t,
\end{equation}
where $\alpha$ is the drift rate (the deterministic trend per unit time) and $\sigma$ is the volatility (standard deviaton of random fluctuations). The increments of this process, denoted by $\Delta x=x_{t+\Delta t}-x_t$ are normally distributed
$$\Delta x \sim N(\alpha \Delta t, \sigma^2 \Delta t).$$

This framework underlier much of modern quantitative finance, including option pricing and risk modeling, as it captures both the continous random variation and systematic trends observed in asset prices.

\subsection{The dynamics of asset log-returns}
Although pioneering, Bachelier model (Eq. \ref{eq:difusaoS}) implied a non-zero probability of negative prices. Theoretical developments driven by the work of \cite{osborne1959brownian} and \cite{samuelson2016proof} established that the diffusion model should be applied to the logarithm of the asset price rather than the price itself.

In financial practice, the analysis of asset time series focuses on returns, normalising price variations. The logarithmic (or continuously compounded) return between time periods $t-1$ and $t$ is defined as follows:
\begin{equation}\label{eq:retorno}
	r_t=\ln(S_t)-\ln(S_{t-1})=\ln\left(\frac{S_t}{S_{t-1}}\right).
\end{equation}
This definition is mathematically convenient and, for small variations, $r_t$ is approximately equal to the simple return $R_t=(S_t-S_{t-1})/S_{t-1}$. The main advantage is that, while the price $S_t$ s constrained to be positive, its logarithm $x_t=\ln(S_t)$ can take any real value, making it an ideal candidate for modelling by a generalised Wiener process.

Therefore, the fundamental postulate of quantitative finance is that the dynamics of the logarithm of an asset's price, $x_t = \ln(S_t)$, is described by:
\begin{equation}
	d\ln(S_t)=v dt + \sigma dX_t.
\end{equation}
This model, known as geometric Brownian motion, captures the two essential characteristics of an asset's dynamics: a continuous expected rate of return, represented by drift $v$ and a random shock term $\sigma X_t$, representing market uncertainty or risk. It forms the basis of the Black-Scholes-Merton model \cite{merton1971theory, black1973pricing} and is the starting point for most pricing models. Against this backdrop, we will present our advection-diffusion model.

\subsection{Diffusion with Retention}\label{dif_retention}

Consider the process of fluid displacement through a certain pipeline. While the dynamics of particle displacement are well modelled by the diffusion process, any obstruction to their movement is disregarded in classical diffusion modelling. Diffusion with retention takes into account external factors that may cause retention of the fluid particles during displacement, such as collision with the pipeline structure, which induces retention of the fluid.

To address the temporary retention effect, additional terms are introduced into the fundamental diffusion equation or the diffusion coefficient is expanded to incorporate higher-order terms \cite{bevilacqua2011significance}.

\subsubsection{Symmetric diffusion with retention}\label{dif_retention}
\begin{figure}[h]
	\center
	\scalebox{0.75}{$
		\begin{tikzpicture}[
			cell/.style={draw, minimum width=1.2cm, minimum height=1.5cm, align=center},
			arr/.style={-{Latex[length=2mm]}, thick}
			]
			\centering
			\node[cell] (t-3) {};
			\node[cell, right=0pt of t-3] (t-2) {};
			\node[cell, right=0pt of t-2] (t-1) {};
			\node[cell, right=0pt of t-1] (t) {};
			\node[cell, right=0pt of t] (t+1) {};
			\node[cell, right=0pt of t+1] (t+2) {};
			\node[cell, right=0pt of t+2] (t+3) {};
			\node[left=10pt of t-3, rotate=90] {$t$};
			
			\foreach \x in {t}{
				\begin{scope}
					\clip (\x.south west) rectangle (\x.north east);
					\fill[gray!40] (\x.south west) rectangle ([yshift=1.5 cm]\x.south east);
				\end{scope}
			}
			
			\node at (t-3.south) [yshift=8pt] {$n-3$};
			\node at (t-2.south) [yshift=8pt] {$n-2$};
			\node at (t-1.south) [yshift=8pt] {$n-1$};
			\node at (t.south) [yshift=8pt] {$n$};
			\node at (t+1.south) [yshift=8pt] {$n+1$};
			\node at (t+2.south) [yshift=8pt] {$n+2$};
			\node at (t+3.south) [yshift=8pt] {$n+3$};
			
			\draw[arr] (t.south) ++(-0.3,-0.05) -- ++(-0.7,-0.8);
			\draw[arr] (t.south) -- ++(0,-0.8);
			\draw[arr] (t.south) ++(0.3,-0.05) -- ++(0.7,-0.8);
			
			\node[cell, below=1.5cm of t-3] (t1-3) {};
			\node[cell, right=0pt of t1-3] (t1-2) {};
			\node[cell, right=0pt of t1-2] (t1-1) {};
			\node[cell, right=0pt of t1-1] (t1n) {};
			\node[cell, right=0pt of t1n] (t1+1) {};
			\node[cell, right=0pt of t1+1] (t1+2) {};
			\node[cell, right=0pt of t1+2] (t1+3) {};
			\node[left=10pt of t1-3, rotate=90] {$t+1$};
			
			\foreach \x/\h in {t1-1/0.8,t1n/1,t1+1/0.8}{
				\begin{scope}
					\clip (\x.south west) rectangle (\x.north east);
					\fill[gray!40] (\x.south west) rectangle ([yshift=\h cm]\x.south east);
				\end{scope}
			}
			
			\node at (t1-3.south) [yshift=8pt] {$n-3$};
			\node at (t1-2.south) [yshift=8pt] {$n-2$};
			\node at (t1-1.south) [yshift=8pt] {$n-1$};
			\node at (t1n.south) [yshift=8pt] {$n$};
			\node at (t1+1.south) [yshift=8pt] {$n+1$};
			\node at (t1+2.south) [yshift=8pt] {$n+2$};
			\node at (t1+3.south) [yshift=8pt] {$n+3$};
			
			\node at (t1-1.north) [yshift=-6pt] {$\gamma$};
			\node at (t1n.north) [yshift=-6pt] {$k$};
			\node at (t1+1.north) [yshift=-6pt] {$\gamma$};
			
			\foreach \x in {t1-1,t1n,t1+1}{
				\draw[arr] (\x.south) ++(-0.3,-0.05) -- ++(-0.6,-0.8);
				\draw[arr] (\x.south) -- ++(0,-0.8);
				\draw[arr] (\x.south) ++(0.3,-0.05) -- ++(0.6,-0.8);
			}
			
			\node[cell, below=1.5cm of t1-3] (t2-3) {};
			\node[cell, right=0pt of t2-3] (t2-2) {};
			\node[cell, right=0pt of t2-2] (t2-1) {};
			\node[cell, right=0pt of t2-1] (t2n) {};
			\node[cell, right=0pt of t2n] (t2+1) {};
			\node[cell, right=0pt of t2+1] (t2+2) {};
			\node[cell, right=0pt of t2+2] (t2+3) {};
			\node[left=10pt of t2-3, rotate=90] {$t+2$};
			
			\foreach \x/\h in {t2-2/0.6,t2-1/0.8,t2n/0.95,t2+1/0.8,t2+2/0.6}{
				\begin{scope}
					\clip (\x.south west) rectangle (\x.north east);
					\fill[gray!40] (\x.south west) rectangle ([yshift=\h cm]\x.south east);
				\end{scope}
			}
			
			\node at (t2-3.south) [yshift=8pt] {$n-3$};
			\node at (t2-2.south) [yshift=8pt] {$n-2$};
			\node at (t2-1.south) [yshift=8pt] {$n-1$};
			\node at (t2n.south) [yshift=8pt] {$n$};
			\node at (t2+1.south) [yshift=8pt] {$n+1$};
			\node at (t2+2.south) [yshift=8pt] {$n+2$};
			\node at (t2+3.south) [yshift=8pt] {$n+3$};

			\node at (t2-2.north) [yshift=-6pt] {$\gamma$};
			\node at (t2-1.north) [yshift=-6pt] {$k \mbox{ } \gamma$};
			\node at (t2n.north) [yshift=-6pt] {$\gamma \mbox{ } k \mbox{ } \gamma$};
			\node at (t2+1.north) [yshift=-6pt] {$\gamma \mbox{ } k$};
			\node at (t2+2.north) [yshift=-6pt] {$\gamma$};
			
		\end{tikzpicture}
		$}
	\caption{Symmetric diffusion with retention, $\gamma=(1-k)/2$.}
	\label{fig-1}
\end{figure}
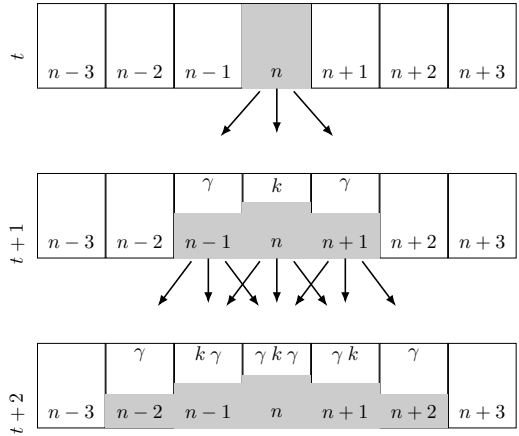

Consider the process described in Figure \ref{fig-1}. The law governing the redistribution of the mass contents of the $n^{th}$ cell, denoted by $p_n$, indicates that a portion of the content, represented by $kp_n$, with $k$ representing the mass fraction, is retained in the n-th cell, while the excess $\gamma p_n$, $\gamma = (1-k)/2$, is uniformly transferred to the neighboring cells at each unit of time. It is noteworthy that the distribution varies over time. If we set $k=0$, we reduce the problem to the Gaussian distribution. Therefore, we have:
\begin{subequations}\label{eq:pnt1epnt2}
	\begin{equation}\label{eq:pnt1}
		p_{n}^{t}=k p_{n}^{t-1}+\gamma p_{n-1}^{t-1}+\gamma p_{n+1}^{t-1},
	\end{equation}
	\begin{equation}\label{eq:pnt2}
		p_{n}^{t+1}=k p_{n}^{t}+\gamma p_{n-1}^{t}+\gamma p_{n+1}^{t},
	\end{equation}
\end{subequations}
where $0\leq k \leq 1$. As \cite{bevilacqua2011significance} pointed out, the model's equations must be derived carefully to ensure consistency with the underlying assumptions. A critical step in the validation process involves testing intermediate expressions against their boundary conditions. For the lower limit $(k=0)$, which corresponds to no retention, the solution must converge to the classical Gaussian distribution. Conversely, for the upper limit $(k=1)$, representing full retention, the solution must be stationary, with all cell contents remaining constant over time.

The detailed derivation of all the intermediate steps leading to the expression below can be found in the Appendix of \cite{bevilacqua2011significance} and does not need to be reproduced here. We have that

\begin{equation}\label{eq:deltabevilacqua}
	\frac{\Delta p_{n}^{t+\Delta t}}{\Delta t}\Delta t=\frac{(1-k)}{2}\left(-(\Delta x)^{4}\frac{k}{2}\frac{\Delta^{4}p_n^{t-\Delta t}}{\Delta x}+(x\Delta)^2\left[\frac{\Delta^2p_n}{(\Delta x)^2}+2\frac{O(\Delta x)^3}{(\Delta x)^2}\right]\right)^{t-\Delta t},
\end{equation}
where $\Delta^2p_n$ and $\Delta^4p_n$ represent the second and fourth order differentials, respectively. Let us take:

\begin{equation}
	\frac{(\Delta x)^2}{\Delta t}=\left(\frac{L_0}{m}\right)^2\frac{m^2}{T_0}=\frac{L_{0}^{2}}{T_0}, \mbox{ }\mbox{ }\mbox{ } \frac{(\Delta x)^4}{\Delta t}=\left(\frac{L_1}{\sqrt{m}}\right)^4\frac{m^2}{T_0}=\frac{L_{1}^{4}}{T_0}, \nonumber
\end{equation}
where $L_0$,$L_1$ and $T_0$  are length and time scale factors, respectively, $\Delta x=L_0/m$ and $\Delta t=L_1/m$ represent the cell size and the time interval, respectively. Substituting the above relations in Eq. \ref{eq:deltabevilacqua}, we get:
\begin{equation}\label{eq:delta4bevilacqua}
	\frac{\Delta p_{n}^{t+\Delta t}}{\Delta t}=\frac{(1-k)}{2}\left\{\frac{L_{0}^{2}}{T_0}\left[\frac{\Delta^2 p_n}{\Delta x^2}+2\frac{O(\Delta x)^3}{\Delta x^2}\right]-\frac{k}{2}\frac{L_{1}^{4}}{T_0}\frac{\Delta^4 p_n}{\Delta x^4}\right\}^{t-\Delta t}.
\end{equation}
The scale factors $L_0$ , $L_1$ and $T_0$ together with the parameter $m$ provide very useful clues to define the sizes of space increment and time step for numerical integration of the finite difference equation.

Note that with $k = 0$, Equation (\ref{eq:delta4bevilacqua}) reduces to the classical diffusion
problem, that is no retention, and with $k = 1$, Equation (\ref{eq:delta4bevilacqua}) represents a
stationary behavior, for the right hand side term of (\ref{eq:delta4bevilacqua}) vanishes. Consequently, the time rate of the contents variation equals zero for all $t$ for all the cells.

By defining $K_2=L_{0}^{2}/2T_0$, $K_4=L_{1}^{4}/4T_0$ and assuming that $p(x,t)$ is a sufficiently well-behaved function of $x$ and $t$, we can take the limits $\Delta x \rightarrow 0$ and $\Delta t \rightarrow 0$ to obtain:

\begin{equation}\label{eq:eqquarta}
	\frac{\partial p}{\partial t}=(1-k)K_2 \frac{\partial^2p}{\partial x^2}-k(1-k)K_4\frac{\partial^4p}{\partial x^4}.
\end{equation}
The fourth-order derivative term introduces the retention effect, where $K_2$ (diffusion coefficient) and $K_4$ (retention coefficient) are generalized constants, $(1-k)$ and $k(1-k)$ are control parameters that express the balance between diffusion and retention when both are simultaneously active. \cite{lipton2002masterclass} and others have shown that stochastic or local volatility models can lead to higher-order partial differential equations (sometimes called ``super-diffusion'').

\subsubsection{Asymmetric diffusion without retention}
Suppose the content within a given cell migrates to neighboring cells according to an asymmetric rule. In this case, we assume no retention. That is
\begin{subequations}\label{eq2:pnt1epnt2}
	\begin{equation}\label{eq2:pnt1}
		p_{n}^{t}=\frac{1}{2}(1+k)p_{n-1}^{t-1}+\frac{1}{2}(1-k)p_{n+1}^{t-1},
	\end{equation}
	\begin{equation}\label{eq2:pnt2}
		p_{n}^{t+1}=\frac{1}{2}(1+k)p_{n-1}^{t}+\frac{1}{2}(1-k)p_{n+1}^{t},
	\end{equation}
\end{subequations}
where $-1 < k < 1$. For $k=0$ this problem reduces to the classical diffusion equation. After some algebraic operations can be found in the Appendix of \cite{bevilacqua2011significance}, we obtained
\begin{equation}\label{assimetric_difusion_ret}
	\frac{\Delta p_{n}^{t+\Delta t}}{\Delta t}\delta t = \Delta x^2 \left\{\frac{(1-k^2)}{2}\frac{\Delta^2 p_n}{\Delta x^2}+\frac{O(\Delta x)^3}{\Delta x^2}\right\}^{t-\Delta t}-\Delta x\left\{k\frac{\Delta p_n}{\Delta x}+\frac{O(\Delta x)^2}{\Delta x}\right\}^{t - \Delta t}.
\end{equation}
Taking
\begin{equation}
	\frac{\Delta x}{\Delta t}=\frac{L_0}{m}\frac{m}{T_0}=\frac{L_0}{T_0}, \mbox{ }\mbox{ }\mbox{ }  \frac{(\Delta x)^2}{\Delta t}=\left(\frac{L_1}{\sqrt{m}}\right)^2\frac{m}{T_0}=\frac{L_1^2}{T_0}, \nonumber
\end{equation}
where $L_0$,$L_1$ and $T_0$  are length and time scale factors, $\Delta x=L_0/m$ and $\Delta t=L_1/m$ represent the cell size and the time interval, respectively. Introducing these expressions in Eq. \ref{assimetric_difusion_ret} and talking the limit when $\Delta x \rightarrow 0$ and $\Delta t \rightarrow 0$, we have:
\begin{equation}
	\frac{\partial p}{\partial t}=\frac{L_1^2}{2T_0}(1-k^2)\frac{\partial^2 p}{\partial x^2}-\frac{L_0}{T_0}k\frac{\partial p}{\partial x}. \nonumber
\end{equation}
Call $K_2=L_1^2/2T_0$ and $K_1=L_0/T_0$ we obtain:
\begin{equation}\label{assimetric_bevilacqua}
	\frac{\partial p}{\partial t}=K_2(1-k^2)\frac{\partial^2 p}{\partial x^2}-K_1k\frac{\partial p}{\partial x}.
\end{equation}
As expected from the distribution rules established for the discrete formulation, the previous equation recovers the classical diffusion-advection equation. In this context, the finite-difference method ensures that the control parameter $k$ appears explicitly in the governing equation, thus preserving its original role. This parameter regulates the rate at which particles are redistributed, maintaining its interpretation as the relative weight between diffusion and propagation. When $k=\pm 1$, propagation dominates preferentially to the right or left, respectively, whereas for $k=0$, the process reduces to purely symmetric diffusion. In the present formulation, the velocity of the advective flow superimposed on the diffusion dynamics coincides with $k$, reflecting the inherent imbalance in the redistribution mechanism \cite{bevilacqua2011significance}.
\section{Methodology}
\subsection{Dynamic of the distribution of asset prices based on three-step random walk}
In this section, we develop our model from first principles. We start with a three-state random walk to model an investor's decision to buy, sell or hold an asset. We show that this discrete process converges to an advection-diffusion equation in the continuous limit. The solution to this equation provides a behavioural foundation for the drift and volatility parameters of the canonical geometric Brownian motion.

It is widely accepted that the logarithm of stock prices follows Brownian motion, meaning that the difference in prices over two discrete time intervals is normally distributed, with the variance proportional to the time interval. The evolution of the probability distribution of prices can be investigated in a manner similar to that employed in \cite{bevilacqua2011significance}. 

Let $S$ be a random variable representing the value of a asset, and let $p(x,t)$ be the density of $x= \log S$ at the point $s$ and at the time $t$. Consider that an individual is willing to pay $S_0$ for a given stock at time $t_0$. At each moment, this individual's preference may remain the same (with probability $k$) or change. If the preference indeed changes, they may be willing to pay $S_{t-1} \Delta s$ with probability $\alpha$ or $S_{t-1}/\Delta s$ with probability $\beta$, such that $\alpha + \beta + k = 1$.

If we consider a large number of individuals and define $p_{N}(x,t)$ as the proportion of individuals, among $N$, who are willing to pay $x$ at time $t$, it is reasonable to asset that $p^*(x,t)=\lim_{N \rightarrow \infty}p_N(x,t)$ represents the probability density for the price. Let $p(x,t)$ be the probability density function of the logarithm of the price, $\Delta x=\log \Delta s$, $x_i=x_0+i\Delta s$ and $t_i=i\Delta t$. Consider a random walk over the positive reals observed at the times $t_i$ such that
\begin{equation}
	x_i=\left\{\begin{array}{rc}
		x_{i-1}+\Delta x,\mbox{with probability } \beta,\\
		x_{i-1},\mbox{with probability } k,\\
		x_{i-1}-\Delta x, \mbox{with probability } \alpha.
	\end{array}\right. \nonumber
\end{equation}
Thus, by setting $p^{t}_{n}=p(x_n,t)$, the fundamental rule of contents redistribution, similar to the process in the Figure \ref{fig-1}, reads:
\begin{eqnarray}\label{rule_dist}
	p_{n}^{t}=kp_{n}^{t-1}+\alpha p_{n-1}^{t-1}+\beta p_{n+1}^{t-1}.
\end{eqnarray}
Subtracting $p_{n}^{t-1}$ in both sides in Eq. (\ref{rule_dist}), we have that
\begin{eqnarray}\label{dif_t-1}
	p_{n}^{t}-p_{n}^{t-1}&=&(k - 1)p_{n}^{t-1}+\alpha p_{n-1}^{t-1}+\beta p_{n+1}^{t-1} \nonumber \\
	&=&\alpha p_{n-1}^{t-1}+\beta p_{n+1}^{t-1}-(\alpha+\beta)p_{n}^{t-1}, \nonumber 
\end{eqnarray}
where $k = 1-\alpha-\beta$. Having the equalities $\alpha=\left(\frac{\alpha+\beta}{2}-\frac{\beta-\alpha}{2}\right)$ and $\beta=\left(\frac{\alpha+\beta}{2}+\frac{\beta-\alpha}{2}\right)$, it follows that
\begin{eqnarray}
	p_{n}^{t}-p_{n}^{t-1}&=&\frac{\alpha+\beta}{2}p_{n-1}^{t-1}-\frac{\beta-\alpha}{2}p_{n-1}^{t-1}+\frac{\alpha+\beta}{2}p_{n+1}^{t-1} \nonumber \\
	&+&\frac{\beta-\alpha}{2}p_{n+1}^{t-1}-(\alpha+\beta)p_{n}^{t-1} \nonumber \\
	&=&(\alpha+\beta)(p_{n-1}^{t-1}-2p_{n}^{t-1}+p_{n-1}^{t-1})+\frac{\beta-\alpha}{2}(p_{n+1}^{t-1}-p_{n-1}^{t-1}). \nonumber 
\end{eqnarray}
Since the function $p$ is sufficiently smooth, the second derivate is finite. Thus, we have (for more details see Appendice)
\begin{eqnarray}
	p_{n}^{t}-p_{n}^{t-1}	&=&(\alpha+\beta)\Delta^2 p_{n}^{t-1}+(\beta-\alpha)\Delta p_{n}^{t-1} + O(\Delta x)^2. \nonumber
\end{eqnarray}
Multiplying both sides by ($\frac{1}{\Delta t}$), having
\begin{eqnarray}
	\frac{p_{n}^{t}-p_{n}^{t-1}}{\Delta t}=\frac{(\Delta x)^2}{\Delta t}(\alpha+\beta)\frac{\Delta^2 p_{n}^{t-1}}{(\Delta x)^2}+\frac{\Delta x}{\Delta t}(\beta-\alpha)\frac{\Delta p_{n}^{t-1}}{\Delta x}+O(\Delta x)^2, \nonumber
\end{eqnarray}
and taking $K_2=\frac{(\Delta x)^2}{\Delta t}$ and $K_1=\frac{\Delta x}{\Delta t}$, in the continous limit, $\Delta x \rightarrow 0$ and $\Delta t \rightarrow 0$, we have that
\begin{eqnarray}
	\frac{\partial p}{\partial t}&=&K_2(\alpha+\beta)\frac{\partial^2 p}{x^2}+K_1(\beta-\alpha)\frac{\partial p}{\partial x} \nonumber \\
	&=&K_2(1-k)\frac{\partial^2 p}{x^2}+K_1(\beta-\alpha)\frac{\partial p}{\partial x}.
\end{eqnarray}
Note that if the inertia probability $k$ is high, volatility decreases, $(1-k)$, that in terms of total trading probability $(\alpha+\beta)$, market volatility is driven by total trading activity. Note that the drift term does not directly depend on $k$. This means that a market with high inertia (many investors sitting idle) can still have a strong trend if the small volume of trading that occurs is strongly imbalanced in one direction.

Now, we solve the PDE to find $p(x,t)$. The previus equation, by taking term relationed of volatility $\mathcal{V} = K_2(\alpha+\beta)$ and the term relationed of the drift $\mathcal{D} = K_1(\alpha-\beta)$, is
\begin{eqnarray*}
	\frac{\partial p}{\partial t}=\mathcal{V}\frac{\partial^2 p}{\partial x^2}-\mathcal{D}\frac{\partial p}{\partial x}.
\end{eqnarray*}
To solve this problem, we use a transformation of variables designed to eliminate the first-order term (advection). Let us set $p=ve^{ax+bt}$. We have that the partial derivatives of $p(x,t)$ are
\begin{eqnarray*}
	\frac{\partial p}{\partial t}=\frac{\partial v}{\partial t}e^{ax+bt}+bve^{ax+bt},
\end{eqnarray*}
\begin{eqnarray*}
	\frac{\partial p}{\partial x}=\frac{\partial v}{\partial x}e^{ax+bt}+ave^{ax+bt},
\end{eqnarray*}
\begin{eqnarray*}
	\frac{\partial^2 p}{\partial x^2}=\frac{\partial^2 v}{\partial x^2}e^{ax+bt}+2a\frac{\partial v}{\partial x}e^{ax+bt}+a^2ve^{ax+bt}.
\end{eqnarray*}
Thus, substituting the values, the equation takes the form
\begin{eqnarray*}
	\frac{\partial v}{\partial t}e^{ax+bt}+bve^{ax+bt}&=&\mathcal{V}\frac{\partial^2 v}{\partial x^2}e^{ax+bt}+2\mathcal{V}a\frac{\partial v}{\partial x}e^{ax+bt}+\mathcal{V}a^2ve^{ax+bt}\\
	&-&\mathcal{D}\frac{\partial v}{\partial x}e^{ax+bt}-\mathcal{D} a v e^{ax+bt},\\
	\frac{\partial v}{\partial t}+bv&=&\mathcal{V}\frac{\partial^2 v}{\partial x^2}+2\mathcal{V}a\frac{\partial v}{\partial x}+\mathcal{V} a^2 v -\mathcal{D}\frac{\partial v}{\partial x}-\mathcal{D}av\\
	\frac{\partial v}{\partial t}&=&\mathcal{V}\frac{\partial^2 v}{\partial x^2}+(2\mathcal{V}a-\mathcal{D})\frac{\partial v}{\partial x}+(\mathcal{V} a^2-\mathcal{D}a-b)v.
\end{eqnarray*}
The purpose of the transformation is to simplify this equation. We choose the constants $a$ and $b$ to cancel the coefficients of the terms $\frac{\partial v}{\partial x}$ and $v$. Let $a$ and $b$ be such that
\begin{eqnarray*}
	\left\{\begin{array}{rc}
		2\mathcal{V}a-\mathcal{D}=0,\\
		\mathcal{V} a^2-\mathcal{D}a-b=0 ,
	\end{array}\right.
\end{eqnarray*}
which gives $a=\mathcal{D}/2\mathcal{V}$ and $b=-\mathcal{D}^2/4\mathcal{V}$. Thus, what remains is to solve the equation
\begin{eqnarray*}
	\frac{\partial v}{\partial t}=\mathcal{V}\frac{\partial^2 v}{\partial x^2},
\end{eqnarray*}
with the initial condition $v(x,0)=\delta(x)$, where $\delta(x)$ is the Dirac delta function, meaning that initially the asset is valued at $0$. However, it is known that the solution to this equation is
\begin{eqnarray*}
	v(x,t)=\frac{1}{\sqrt{4\pi \mathcal{V}t}}exp\left(-\frac{x^2}{4\mathcal{V}t}\right).
\end{eqnarray*}
This provides us with
\begin{eqnarray*}
	p(x,t)&=&\frac{1}{\sqrt{4\pi A^2t}}\exp\left(-\frac{x^2}{4\mathcal{V}t}\right)\exp\left(\frac{\mathcal{D}x}{2\mathcal{V}}-\frac{\mathcal{D}^2t}{4\mathcal{V}}\right)\\
	&=&\frac{1}{\sqrt{4\pi \mathcal{V}t}}\exp\left(\frac{-x^2+2\mathcal{D}tx-\mathcal{D}^2t^2}{4\mathcal{V}t}\right)\\
	&=&\frac{1}{\sqrt{4\pi \mathcal{V}t}}\exp\left(-\frac{(x-\mathcal{D}t)^2}{4\mathcal{V}t}\right).
\end{eqnarray*}
This is the probability density function of a normal distribution, $N(\mathcal{D}t, 2\mathcal{V}t)$. The parameter D directly governs the mean of the distribution, $E[X_t] = \mathcal{D}_t$. This establishes it as the drift rate, or the continuous expected return of the logarithm of the price. This parameter is a direct consequence of the asymmetry between trading probabilities. If the probability of $\alpha$ exceeds that of $\beta$, then $\mathcal{D}$ is positive, generating a downward trend. If $\beta > \alpha$, the trend is upward. Thus, the model provides a behavioural foundation for market drift.

The parameter $\mathcal{V}$ determines the variance of the distribution, $Var[X_t]=2\mathcal{V}t$. It represents the diffusion coefficient and is the fundamental source of uncertainty and risk in the model. A higher value of $\mathcal{V}$ implies greater dispersion of possible price outcomes, which broadens the distribution curve and increases the probability of extreme price movements. The variance grows linearly with time, which is a characteristic of Brownian motion that is also present in our model.

The result found allows a direct bridge with the standard model of finance, where the log-price also follows a Generalized Brownian Motion, $dx_t=vdt+\sigma d X_t$, whose solution is $N(vt, \sigma^2t)$. By comparing the two solutions, we can establish a direct correspondence between the parameters:
\begin{itemize}
	\item[i.] Mean: $\mathcal{D}t = vt \Rightarrow \mathcal{D}=v$;
	\item[ii.] Variance: $2\mathcal{V}t = \sigma^2 t \Rightarrow \sigma^2=2\mathcal{V}$.
\end{itemize}
The most important relationship here is $\sigma=\sqrt{2\mathcal{V}}$. This means that volatility $\sigma$, one of the most crucial parameters in finance, is directly linked to our diffusion coefficient $A$. Our model therefore offers both a physical and a microeconomic interpretation for the origin of market volatility.

\subsection{Modified diffusion with retention for t-Student model}
The three-step random walk model has a significant advantage in that it derives the market's drift and volatility from the microeconomic principles of buying, selling and holding decisions. This establishes a clear interpretative link between agent behaviour and the classic advection-diffusion equation. However, the resulting Gaussian distribution fails to capture the leptokurtosis, or 'fat tails', that are characteristic of financial returns. This limitation is inherent in all models based on standard Brownian motion.

To address this empirical shortcoming, we turn to the diffusion-with-retention framework proposed by \cite{bevilacqua2011significance}. Notably, \cite{bevilacqua2015second} reinforced the relevance of this physical framework for economics in a subsequent paper. In that work, he presents a conceptual application of this 'second-order diffusion theory' to model capital flow. He argues that its ability to handle two simultaneous opposing fluxes (``bi-flux'') makes it better suited to describing economic phenomena such as income and expenditure.

Building on this precedent, we propose the following conceptual modification to asset pricing: the parameter $k$, for which $0 \leq k \leq 1$, originally representing physical retention, can be reinterpreted as a more sophisticated measure of investor inertia. In this view, inertia dampens overall trading activity and fundamentally alters the statistical nature of the price diffusion process by introducing higher-order differential terms. The governing equation for the probability density $p(x, t)$, similary to as seen in Section \ref{dif_retention}, is given by
$$\frac{\partial p}{\partial t}=(1-k)K_2 \frac{\partial^2p}{\partial x^2}+k(1-k)K_4\frac{\partial^4p}{\partial x^4}.$$
Note that, we modified se signal with sum the four order term. The key insight is the emergence of the fourth-order derivative term, $\frac{\partial^4p}{\partial x^4}$. In financial modelling, these terms are known to generate distributions with a higher degree of kurtosis than the normal distribution. The coefficient $k(1-k)K_4$ directly controls the magnitude of this effect. When inertia $k$ is zero, the term disappears and the classic diffusion equation is recovered. As inertia increases, however, the fourth-order term becomes more prominent. This causes the probability distribution to develop a sharper peak and heavier tails. Thus, it provides a potential endogenous explanation for the leptokurtosis observed in the market.

This extension enables us to model a market in which periods of indecision $(k > 0)$ reduce overall volatility and increase the likelihood of rare, significant price fluctuations. This makes the model more aligned with empirical reality.

The probability density function (PDF) of the fourth-order partial differential equation Eq. \eqref{eq:eqquarta} does not have a trivial analytical solution. However, the mathematical effect of the fourth-order term $\frac{\partial^4p}{\partial x^4}$ is well understood: it shifts probability mass from the 'shoulders' of the distribution to the centre and the 'tails'. This results in a distribution with a higher peak (leptokurtic) \cite{tsallis2009introduction} and heavier tails than the normal distribution.

The t-Student's distribution is a well-established parametric distribution which exhibits these characteristics precisely. The degrees of freedom parameter, $v$, directly controls the distribution's kurtosis. For low $v$, the Student's t-distribution has very heavy tails, and as $v$ tends to infinity, it converges to the normal distribution. Therefore, rather than solving the complex partial differential equation (PDE) numerically, we use the Student's t-distribution as a phenomenological approximation or an effective proxy. This provides a tractable and well-understood tool that captures the key statistical signature of heavy tails that the fourth-order term aims to produce. Modelling the distribution of stock returns using the Student's t-distribution is a classic approach in the literature \cite{praetz1972distribution}, justified by its superior ability to fit empirical data compared to the normal distribution.

\subsubsection{Derivation of variance and kustosis associated with t-Studant model}
In the PDE with four order term, the term $\frac{\partial^2 p}{\partial x^2}$ is the standart diffusio term associated with variance and four order term $\frac{\partial^4 p}{\partial x^4}$, is the term that introduces the leptokurtosis in the model \cite{tsallis2009introduction}.

Define the $n$-th moment how
\begin{align*}
	x^{(n)}(t)=\int_{-\infty}^{\infty}x^n p(x,t) dx.
\end{align*}
To find the temporal evolution of $x^{(n)}(t)$, we derive with respect to $t$:
\begin{align*}
	\frac{d}{d t}x^{(n)}(t)=\int_{-\infty}^{\infty}x^n \frac{\partial p}{\partial t} dx.
\end{align*}
Thus, we have
\begin{align}
	\frac{d}{d t}x^{(n)}(t)=\int_{-\infty}^{\infty}x^n\left\{(1-k)K_2 \frac{\partial^2p}{\partial x^2}+k(1-k)K_4\frac{\partial^4p}{\partial x^4}\right\}dx
\end{align}
Now, we use piecewise integration repeatedly. For a term with order $j$, the general identity for piecewise integration is
\begin{align*}
	\int_{-\infty}^{\infty}x^n \frac{\partial^j p}{\partial x^j}dx=\left[\sum_{i=0}^{j-1}(-1)^i\frac{d^i (x^n)}{d x^i}\frac{\partial ^{j-1-i}p}{\partial x^4}\right]_{-\infty}^{\infty}+(-1)^i\int_{-\infty}^{\infty}\frac{d^j(x^n)}{d x^j}p(x,t)dx.
\end{align*}
Assuming coutorn conditions \cite{risken1989fokker}, that is, $p(x,t)$ and its derivatives go to zero at $\infty$, we have
\begin{align*}
	\lim_{|x| \rightarrow \infty}x^m \frac{\partial^r p}{\partial x^r}=0,
\end{align*}
for all $m < n$ and results derivates of order $r$. This condition implies that $p(x,t)$ must decay more fast than any polynomial $x^n$ it grows into infinity. Under this assumption, the integrals is simplified (for more details see Appendix), for $j=2$, applying integration piecewise
\begin{align}
	\int x^n \frac{\partial^2 p}{\partial x^2}dx=n(n-1)x^{(n-2)},
\end{align} 
and for $j=4$
\begin{align}
	\int x^n\frac{\partial^4 p}{\partial x^4}dx=n(n-1)(n-2)(n-3)x^{(n-4)}.
\end{align}
Assuming that $x^{(1)}=0$, we apply the equations above for $n=2$
\begin{align*}
	\frac{d^2}{d t}=(1-k)K_2 2 x^{(0)} \Rightarrow \frac{d}{dt} Var(t)=2(1-k)K_2 t.
\end{align*} 
From this, with initial condition $Var(0)=0$, we have, integrating with respect to $t$
\begin{align}
	Var(t)=2(1-k)K_2 t.
\end{align} 

For the four order term, we apply, for $n=4$
\begin{align*}
	\frac{d}{d t}x^{(4)}&=(1-k)K_2 12x^{(2)}+k(1-k)K_4 24x^{(0)}\\
	&=24(1-k)^2 K_{2}^{2} t+24k(1-k)K_4.
\end{align*}
From this, integrating with respect to $t$, we have
\begin{align*}
	x^{(4)}(t)=12(1-k)^2 K_{2}^{2}t^2+24 k(1-k)K_4 t.
\end{align*}
Thus, the kurtosis, defined by $\kappa (t)=\frac{x^{(4)}}{(x^{(2)})^2}$, is give by
\begin{align}
	\kappa (t)=\frac{12(1-k)^2 K_{2}^{2}t^2+24 k(1-k)K_4 t}{4(1-k)^2K_2^2 t^2}=3+\frac{6 k K_4}{(1-k)K_{2}^{2}t},
\end{align}
and the excess kurtosis is give by
\begin{align*}
	\kappa_e (t) = \frac{6 k K_4}{(1-k)K_{2}^{2}t}.
\end{align*}
\section{Simulation and Empirical Application for Models Validation with Market Data}
To validate the robustness and applicability of the proposed models, we ran a simulation and completed a practical exercise\footnote{The complete R code used for data analysis, model calibration, simulation, and the generation of all figures presented in this paper is publicly available in a GitHub repository: https://github.com/DiegoSsant/Financial-Modeling-Diffusion-Retention}. The aim was to compare the performance of the advection-diffusion model (derived from our random walk model) with that of the extended diffusion model with retention, using historical data from a real asset as a reference. This procedure determines whether, once calibrated, the models can generate price trajectories and return distributions that are statistically consistent with observed market dynamics.

The asset selected for analysis was the preferred stock of Petrobras (PETR4.SA), one of the most liquid securities traded on the Brazilian stock exchange, B3. Adjusted closing prices were collected for the period from January 1, 2020, to July 31, 2025, using the Yahoo Finance platform\footnote{Data sourced from Yahoo Finance, available at https://finance.yahoo.com/quote/PETR4.SA.}. Based on these prices, daily log-returns were computed as $r_t = \ln(S_t/S_{t-1})$, which serve as the empirical counterpart to the infinitesimal increments $dx$ of our continuous-time process.

\subsection{Model calibration}
\subsubsection{Advection-diffusion model}
The advection-diffusion model parameters, $\mathcal{D}$ (drift) and $\mathcal{V}$ (diffusion), were estimated directly from the log-return series. Assuming a time step $\Delta t = 1$ for daily data, the parameters were obtained as follows:
\begin{itemize}
	\item The drift parameter $\mathcal{D}$ was set equal to the sample mean $\bar{x}$ of the log-returns, aligning the expected trend of the model with the historical daily average return of the asset;
	\item The diffusion parameter $\mathcal{V}$ was computed as half the sample variance of the log-returns, $\mathcal{V} = s^2/2$.
\end{itemize}
These are the maximum likelihood estimators of normal distribution. With $\mathcal{V}$ and $\mathcal{D}$ estimated, a synthetic price trajectory was simulated starting from an initial price of 100. For each trading day, a random increment was drawn from a normal distribution with mean $\mathcal{D}\Delta t$ and variance $2\mathcal{V}\Delta t$. The log price was updated additively and then exponentiated to recover the price level. To enable a direct visual comparison, the historical price series was normalised to start at the same initial value of 100.

\subsubsection{Diffusion with retetion model}
For the diffusion with retention model, we derive the formulas for variance and kurtosis, which are the most important parameters for this model as they are most influenced by the terms $k$, $K_2$ and $K_4$. The model should generate these formulas for a small time step, $\Delta t=1$, which in our case is equal to one day.
\begin{itemize}
	\item The variance is primarily controlled by the second-order term. The formula is as follows:
	\begin{equation*}
		\sigma^2=2(1-k)K_2\Delta t.
	\end{equation*}
	\item Kurtosis, or 'heavy tails', is generated by the interaction between second- and fourth-order terms. The approximate formula for excess kurtosis is
	\begin{equation*}
		\kappa_e = \frac{6 k K_4}{(1-k)K_{2}^{2}\Delta t}.
	\end{equation*}
\end{itemize}
The retention model lacks a simple analytical solution and was therefore calibrated using the Moment Calibration method. The process consisted of three steps:
\begin{itemize}
	\item[1.] Determining Empirical Targets: First, we calculate the crucial moments of PETR4.SA log-return series: the empirical variance $\sigma_{emp}^{2}$ and the empirical excess kurtosis $\kappa_{emp}^{2}$. The values serve as the numerical targets that the model should replicate.
	\item[2.] Defining the objective function: Next, we define an objective function that measures the difference between the theoretical moments of the model and the empirical targets. To achieve this, we use theoretical formulas for variance $\sigma^2$ and kurtosis $\kappa$. The objective function calculates the sum of squared errors between theoretical and empirical values.
	\item[3.]  Numerical optimisation: Finally, we use a numerical optimisation algorithm to determine the optimal combination of parameters $(k, K_2, K_3)$ that minimizes the objective function. The result is calibrated parameters that best describe observed return dynamics from the perspective of the retention model. 
\end{itemize}

\subsection{Simulation Framework}
Once the models had been calibrated, multiple price trajectories were simulated for each model in order to compare their behaviour with historical data. To enable a robust visual comparison, five simulations were generated for each model.

In the case of the advection-diffusion model, daily random increments were drawn from a normal distribution based on the estimated $\mathcal{D}$ and $\mathcal{V}$ parameters.

For the diffusion model with retention, we used a Student's t-distribution as a proxy for the simulation because it can generate the heavy tails (leptokurtosis) produced by the fourth-order equation. The shape of the Student's t-distribution is controlled by the $v$ parameter representing the degrees of freedom, which was adjusted to ensure that the kurtosis of the simulation matched that of reality perfectly. To this end, the df value was calculated directly from the empirical excess kurtosis $\kappa_{emp}$ using the following relationship $v=6/\kappa_{emp}+4$.

\subsection{Results and Analysis}
To analyse the dynamic behaviour of the models, five price trajectories were generated for each one and compared with the normalized historical price of PETR4.SA (black line).

\begin{figure}[!htp]
	\centering
	\includegraphics[width=0.75\linewidth]{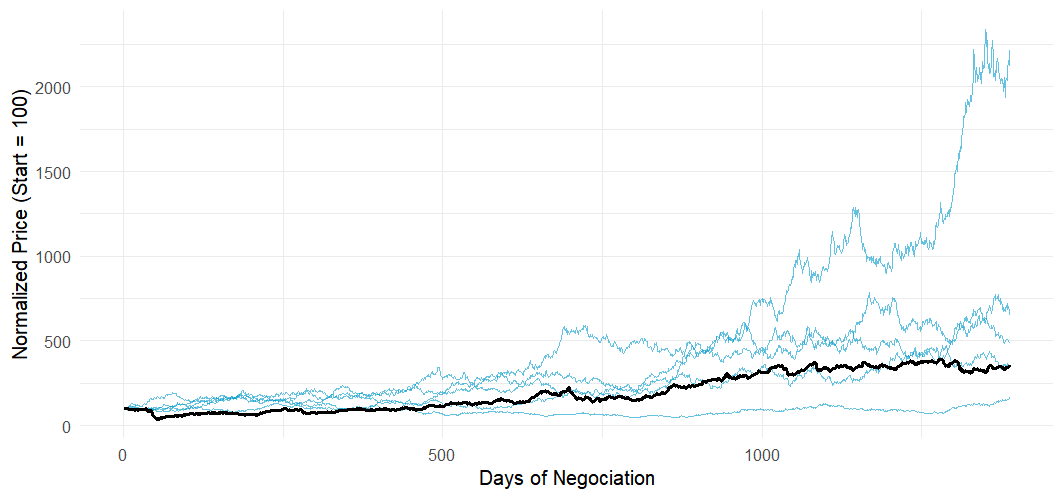}
	\caption{Comparison of the asset's real price of PETR4.SA vs. simulated price (Normalized).}
	\label{fig:simu}
\end{figure}
\begin{figure}[!htp]
	\centering
	\includegraphics[width=0.75\linewidth]{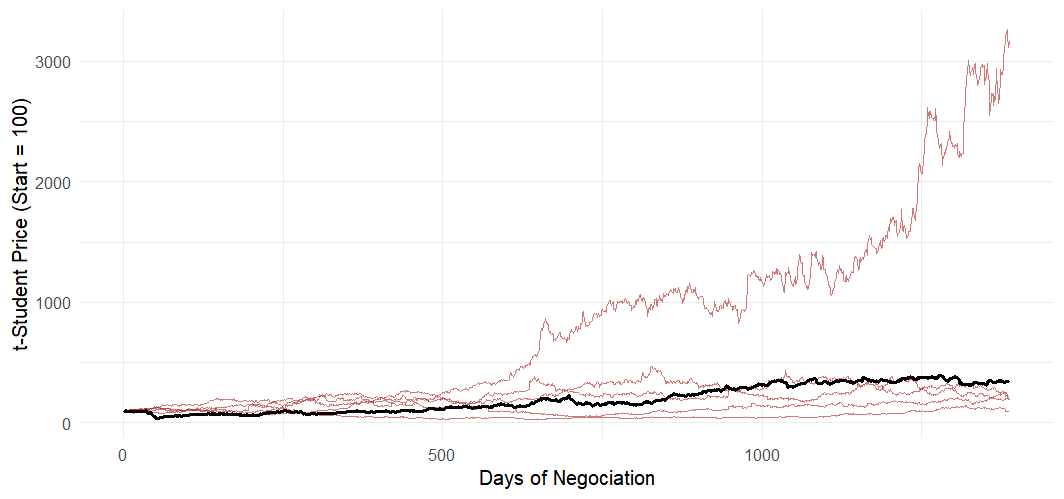}
	\caption{Comparison of the asset's real price of PETR4.SA vs. simulated price (t-Student).}
	\label{fig:simu2}
\end{figure}
Figure \ref{fig:simu} illustrates the possibilities generated by the advection-diffusion model, as shown by the simulations of the Normal model. There is a very large dispersion among the simulated trajectories (blue lines). While some trajectories reach extremely high price levels (above 2000), far beyond historical observations, others remain at lower levels. This demonstrates how, although Gaussian shocks are small on average, they can accumulate and lead to scenarios of significant divergence.

Figure \ref{fig:simu2} shows the trajectories generated by the heavy-tailed model. Although variability remains high, the behaviour of the simulations (red lines) appears more consistent visually. The presence of larger and more frequent shocks, which are characteristic of a Student's t-distribution, appears to induce stronger reversals. This prevents most of the trajectories from deviating dramatically and sustainably in one direction. While one of the trajectories still reaches a high peak, as would be expected of a heavy-tailed model, the set of simulations as a whole appears to generate a more plausible range of possibilities in relation to the behaviour of the real asset.

\begin{figure}[!htp]
	\centering
	\includegraphics[width=0.8\linewidth]{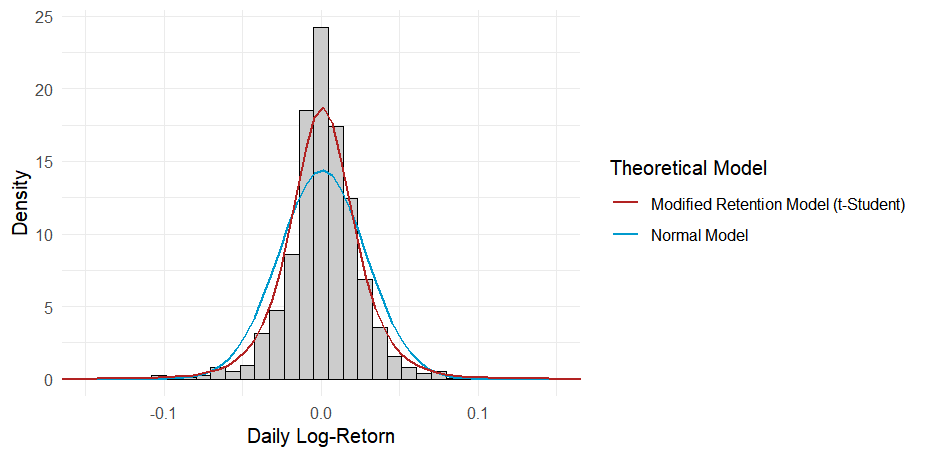}
	\caption{Histogram of Log-Returns of PETR4.SA in which the theoretical normal and t-Studant curves is shown in blue and red versus the actual data.}
	\label{fig:apli}
\end{figure}

The strongest evidence for the superiority of the retention model is its ability to replicate the probability distribution of daily log returns. Figure \ref{fig:apli} plots the theoretical densities of both models on a histogram of PETR4.SA's empirical returns.
The normal model (blue curve), which is derived from the advection-diffusion process, fails to capture the two most striking features of the data: 
\begin{itemize}
	\item Leptokurtosis: The blue curve's peak is clearly lower and wider than the histogram's sharp peak, which underestimates the frequency of days with near-zero returns.
	\item Heavy tails: At the extremes of the distribution, the blue curve approaches zero very quickly. This ignores the probability of extreme events (large gains or losses), which, although rare, are evident in the data.
\end{itemize}

By contrast, the Retention Model (red curve), which is approximated by a calibrated Student's t-distribution, provides a notably superior fit. Its taller, narrower peak aligns perfectly with the central concentration of returns and its heavier tails decay more slowly, providing adequate coverage of the observed extreme returns.

\begin{figure}[!htp]
	\centering
	\includegraphics[width=0.8\linewidth]{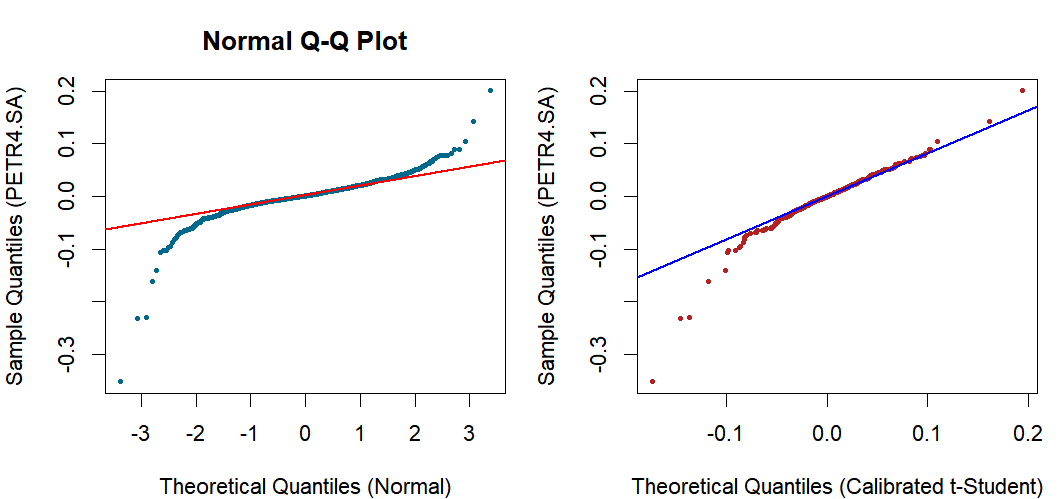}
	\caption{Quantile-Quantile (Q-Q Plot) of Log-Returns of PETR4.SA.}
	\label{fig:apli_2}
\end{figure}

This conclusion is further reinforced by the Q-Q (quantile-quantile) plots in Figure \ref{fig:apli_2}.

The normal Q-Q plot (left) shows the standard deviation in the form of an 'S'. At the tails, the sample quantiles (blue dots) systematically deviate from the theoretical reference line (red), which confirms that the normal model underestimates the magnitude of rare events.

The Retention Q-Q plot on the right shows a dramatic improvement. The sample quantiles (red dots) adhere closely to the theoretical reference line (blue) throughout their entire length, including at the extreme tails. This near-perfect alignment provides strong visual evidence that the retention model successfully captures the quantile dependence structure of the real data.

\section{Conclusion}
This work addressed a fundamental gap in standard asset pricing models by formally incorporating investor inertia. Addressing this issue led to two main contributions from our research, offering a new perspective on how the behaviour of market agents shapes price dynamics.

The first contribution was the derivation of the classical advection-diffusion model from a three-state random walk involving buying, selling or holding. This is a significant result because it provides an intuitive microfoundation for the macroscopic parameters of the market. We demonstrate that drift naturally emerges from the asymmetry between buying and selling decisions, while volatility is a direct consequence of total trading activity. However, this model inherits the well-known inability to replicate the stylised facts of financial markets, particularly leptokurtosis, due to its resulting Gaussian distribution for returns.

To overcome this limitation, we introduced and validated a fourth-order extension based on the diffusion model with retention, which was our second and main contribution. Justified by previous conceptual applications in capital flows, this approach enabled us to reinterpret 'retention' as a more complex manifestation of investor inertia. An empirical analysis conducted using data from the PETR4.SA asset unequivocally validated the superiority of this extended model. Distributional fit analysis (histograms and Q-Q plots) and simulations of multiple price paths confirmed that the retention model remarkably accurately captures the heavy tails and sharp peaks of the return distribution features that the normal model fails to describe.

The most profound implication of our work is the suggestion that investor inertia has a 'dual nature'. In its simplest form, which is the decision not to trade, it acts as a modulator of volatility within a Gaussian paradigm. In its more complex form, represented mathematically by 'holding', it fundamentally alters the nature of the stochastic process, resulting in tail risk and an increased probability of extreme events that characterise real markets.

Although our study successfully applied the fourth-order model, a rigorous derivation of the equation based on behavioural principles rather than a physics analogy, this remains a fertile area for future research. Further extensions could include applying this framework to pricing derivatives or developing more realistic risk metrics, such as Value at Risk (VaR), which incorporate the heavy tails generated by the model endogenously.

\section*{Acknowledgement(s)}

We gratefully acknowledge the financial support provided by the Coordination for the Improvement of Higher Education Personnel (CAPES) under Grant $88887.679920/2022-00$.

\section*{Disclosure statement}
No potential conflict of interest was reported by the author(s).

\section*{Statement on the Use of AI Tools}
During the preparation of this work, the authors used the Gemini 2.5 pro language model (Google) as an aid for the following tasks:
\begin{itemize}
	\item[1.] Reviewing and improving the clarity, grammar, and style of the text.
	\item[2.] Assistance in preparing application code and simulations in R software.
	
\end{itemize}
The authors reviewed, edited, and validated all content and assume full responsibility for the originality, accuracy, and integrity of the final version of the manuscript.

\bibliographystyle{unsrtnat}
\bibliography{references} 
\section*{Appendix}\label{appendix}

\section*{Elementary  Relations  of  Finite  Difference Mathematics}
Let $f:\mathbb{R} \rightarrow \mathbb{R}$ be a function mapping  the  set  of  real  numbers  onto 
itself. The  following  notation  will  be  used  throughout  this  paper: $$f_k \equiv f_k(x) \equiv f(x_k).$$
The m-th order difference of $f_k(x)$ is written as $\Delta^m f_k$. The m-th order difference is the finite difference approximation of the m-th derivate, $\partial^m f(x)/\partial x^m$, of $f(x)$.
Define $O(\Delta x)^j$ as algebraic expressions multiplied by terms of order $(\Delta x)^j$ or higher. For instance: $O(\Delta x)^3 \propto \alpha O(\Delta x)^3+\beta O(\Delta x)^4$, $\alpha < \infty$ and $\beta < \infty$.

If $f(x)$ and all derivates $\partial^m f(x)/\partial x^m, r=1, \dots, j$, are continous, the funcsiont is said to belong the class $C^j$, $f(x) \in C^j$.

The difference of order $m>1$, for $f(x) \in C^j$, is give by
$$\Delta^m f_k = \sum_{i=0}^{m}f_{k+z-i}(-1)^i C_{m}^{i},$$
where $C_{m}^{i}=m!/(m-i)!i!$ and $z=(-m-\{m mod2\})/2$. Examples of $\Delta^m f_k$ are
\begin{eqnarray}
	\Delta f_k &=& f_k-f_{k-1}, \nonumber \\
	\Delta^2 f_k &=& f_{k-1}-2 f_k+f_{k+1}, \nonumber \\
	\Delta^3 f_k &=& f_{k-2}-3 f_{k-1}+3 f_{k}-f_{k+1}. \nonumber
\end{eqnarray}
The Taylor expansion for sufficiently smooth function is:
\begin{eqnarray}
	f_{k+1}=f_{k}+\Delta x\frac{\partial f_{k+1}}{\partial x_k}+O(\Delta)^2. \nonumber
\end{eqnarray}
Since the function $f(x)$ is sufficiently smooth, the second derivate is finite and we many write:
\begin{eqnarray}
	\Delta f_{k+2}=\Delta f_{k+1}+O(\Delta x)^2. \nonumber
\end{eqnarray}
Therefore, the differences centered at two neighboring points differ by terms of order $O(\Delta x)^2$. In general we have:
\begin{eqnarray}
	\Delta^m f_{k+2}=\Delta^m f_{k+1}+O(\Delta x)^{m+1}. \nonumber
\end{eqnarray}
From the expressions derived earlier, one can readily establish another relation that will facilitate the forthcoming analysis: 
\begin{eqnarray}
	f_{k+2}&-&f_{k}=f_{k+2}-f_{k+1}+f_{k+1}-f_{k}=\nonumber \\
	\Delta f_{k+2} &+& \Delta f_{k+1}=2\Delta f_{k+1}+O(\Delta x)^2. \nonumber
\end{eqnarray}
\section*{Obtention of piecewise integrations for the t-Student model's variance and kurtosis derivation}
Under the assumptions coutorn conditions, $\lim_{|x| \rightarrow \infty}x^m\frac{\partial^r p}{\partial x^r}=0$, take the integrals of interest
\begin{align*}
	I_2=\int_{-\infty}^{\infty}x^n \frac{\partial^2 p}{\partial x^2}dx \mbox{ } \mbox{ and } \mbox{ } I_4=\int_{-\infty}^{\infty}x^n \frac{\partial^4 p}{\partial x^4}dx.
\end{align*}
Applying integration piecewise to the integral $I_2$, we have
\begin{align*}
	I_2=\left[x^n\frac{\partial p}{\partial x}\right]_{-\infty}^{\infty}-\int_{-\infty}^{\infty}n x^{n-1}\frac{\partial p}{\partial x}dx ,
\end{align*}
and, by assumption $\left[x^n\frac{\partial p}{\partial x}\right]_{-\infty}^{\infty}=0$. Integrating piecewise again, we have
\begin{align*}
	I_2&=-n\left(\left[x^{n-1}p\right]_{-\infty}^{\infty}-\int_{-\infty}^{\infty}p(n-1)x^{n-2}dx\right)\\
	&=n(n-1)\int_{-\infty}^{\infty}x^{n-2}p(x,t) dx\\
	&=n(n-1)x^{(n-2)}.
\end{align*}
Applying integration piecewise to the integral $I_4$, we have, by assumption
\begin{align*}
	I_4&=\left[x^n\frac{\partial^3 p}{\partial x^3}\right]_{-\infty}^{\infty}-\int_{-\infty}^{\infty}n x^{n-1}\frac{\partial^3 p}{\partial x^3}dx\\
	&=-\int_{-\infty}^{\infty}nx^{n-1}\frac{\partial^3 p}{\partial x^3}dx.
\end{align*}
Applying integration piecewise recursively, we have
\begin{align*}
	I_4=\int_{-\infty}^{\infty}x^n\frac{\partial^4 p(x,t)}{\partial x^4}=n(n-1)(n-2)(n-3)x^{(n-4)}(t),
\end{align*}
having, by the assumption coutorn conditions $\left[x^n\frac{\partial^3 p}{\partial x^3}\right]_{-\infty}^{\infty}=0$, $-\left[nx^{n-1}\frac{\partial^2 p}{\partial x^2}\right]_{-\infty}^{\infty}=0$, $\left[n(n-1)x^{n-2}\frac{\partial p}{\partial x}\right]_{-\infty}^{\infty}=0$ and $-\left[n(n-1)(n-2)x^{n-3}p(x,t)\right]_{-\infty}^{\infty}=0$.
\end{document}